\documentclass[a4paper,twocolumn,10pt,submitted]{quantumarticle}
\pdfoutput=1
\usepackage[utf8]{inputenc}
\usepackage[english]{babel}
\usepackage[T1]{fontenc}
\usepackage{amsmath}
\usepackage{hyperref}

\usepackage{tikz}
\usepackage{lipsum}

\usepackage{enumerate}
\usepackage{graphicx}
\usepackage{color}
\usepackage{enumitem}
\usepackage{appendix}
\usepackage{ulem}  

\usepackage{fontenc}  

\usepackage{tikz}
\usetikzlibrary{positioning, decorations.pathmorphing}
\usetikzlibrary{matrix}

\newcommand{\beq}{\begin{equation}}
\newcommand{\eeq}{\end{equation}}
\newcommand{\bea}{\begin{eqnarray}}
\newcommand{\eea}{\end{eqnarray}}
\newcommand{\ba}{\begin{array}}
\newcommand{\ea}{\end{array}}

\newcommand{\bef}{\begin{figure}}
\newcommand{\eef}{\end{figure}}

\begin{document}

\title{\color{black} Heading towards an Algebraic Heisenberg Cut}

\author{Mathias Van Den Bossche}
\affiliation{Thales Alenia Space, 26, avenue J.-F. Champollion, 31037 Toulouse, France}
\author{Philippe Grangier}
\affiliation{Laboratoire Charles Fabry, IOGS, CNRS, 
Universit\'e Paris Saclay, F91127 Palaiseau, France.}

\begin{abstract}
In previous papers  we have explained how a sequence of theorems by John von Neumann on infinite tensor products (ITP) can be understood as providing elements to support both sectorisation of the Hilbert space of large quantum systems, and a mechanism of self decoherence thereof. These two effects may help understanding the articulation of the classical and quantum realms. However, as they involve considering an infinite number of quantum degrees of freedom, legitimate concerns can be raised on their applicability.  In this paper, we address explicitly the interface between both realms through the example of a simplified model of a photon polarisation measurement device. Guided by the fact that there is von Neumann sectorisation at infinity, and by the necessity of classical contexts to perform measurements, we show that this limit can be under control, and that although the full force of the sectorisation theorems requires taking the infinite limit, early signs of the macroscopic behaviour appear before infinity. In our example, this shows up in photodiodes through diverging electron avalanches that simultaneously make the system classical, localise it randomly in a macroscopic sector and provide a macroscopic signal. This lays the grounds for justifying the inclusion in quantum physics of the ITP formalism, which involves non-separable Hilbert spaces and potentially type-III von Neumann algebras. Such an approach could make sense of the quantum-classical transition as a primarily algebraic one.

\end{abstract}

\maketitle

\section{Motivations}
Since the formalisation of quantum theory almost one century ago, the difference between what can be expected from a system in the quantum realm (if any) and what we are used to in the classical realm has been continuously raising questions at many levels. The core of the differences {\color{black} is} located in the measurement process, that involves a macroscopic measurement device to extract information from the quantum system. We hold as key that this macroscopic aspect is not a spurious detail of the framework, but {\it the} structuring element that makes measurement contextual, and quantum physics intrinsically different from classical physics. {\color{black} This  point of view developed in \cite{CO2002,csm1,csm2,csm4} and summarized in Annex~1  will be pursued in the present article.}

In a series of papers  
\cite{completing,ITP1,ITP2,ITP3} we have argued that it could make sense to investigate what can be obtained from the properties of infinite tensor products (ITP) of $N$ elementary Hilbert spaces,  to describe the  behaviour of quantum systems at the macroscopic limit $N\rightarrow \infty$. 
Of particular interest is { \color{black} a set of `sectorisation' theorems by John von Neumann \cite{JvN39} that explain how ITP Hilbert spaces (i) break down into an uncountable number of orthogonal separable subspaces that we call {\it sectors,} 
{\color{black} defined so that their}
direct sum is the full ITP space\footnote{\color{black} Von Neumann calls them `incomplete direct products', i.e. incomplete tensor product in current terminology. The full (complete) tensor product is not only infinite but non-separable.}   and (ii) that these sectors are not connected by operators built from operators in each elementary Hilbert space. This} implies a form of self-decoherence without tracing on external degrees of freedom. The algebras of operators that are relevant to describe the observables in this limit {\color{black} are not necessarily type-I von Neumann (W*-)algebras, 
as it would be the case for operators on the usual finite or countably-infinite dimension Hilbert spaces. Here, {\color{black} the ITP becomes nonseparable and }up to type-III algebras may be required \cite{completing,ITP1,ITP2,ITP3}}. 
Even though considering the large $N$ limit is common in statistical physics \cite{emch}, it needs to be checked when applied to a new case. {\color{black}This is even more true here since a qualitative change seems to appear only at infinity -- which} has always been a tricky topic to handle. In the present case, it means going beyond Streater and Wightman vulgate assumptions \cite{bookSW,nonsep}, {\color{black} in particular by considering also non-separable Hilbert spaces}, giving to the whole picture a fairly abstract mathematical content. 

In this paper we go back to physics, {\color{black} by considering a model of photon polarisation measurement device, simplified {\color{black} down to a Gedankenexperiment style,} by removing all conventional sources of decoherence. We will focus on a direct, destructive measurement, but the calculation can be easily extended to an indirect, quantum non-demolition one (QND, see Annex~2). 
This  will allow us 
to show how these mathematical considerations could be at the fundamental core of measurement processes,} 
{\color{black} 
shedding light on the subtle articulation between the classical and the quantum worlds, that seem 
to require each other  within a unified physical reality \cite{enigma}.}
\vskip 1mm 

This article is organised the following way. We first define the {\color{black} notion of}  `sector parameter' observable that allows labelling macroscopic states. We then introduce the {\color{black} simplified model that we consider to describe the avalanche photodiode (APD), which} will be at the core of the polarisation measurement device. Next we build the avalanche quantum state, 
explore {\color{black} its large $N$} properties, {\color{black} and analyse their consequence in the frame of the generic measurement model presented in \cite{genModel}}. We then introduce the sector parameter {\color{black} relevant to this case} and we compute the large-$N$ behaviour of its expectation value in the avalanche state. We finally discuss our results, explaining the {\color{black}subtlety} of the limit and highlighting differences with previous attempts at modeling this kind of phenomena 
-- in particular, {\color{black} how }the usual Bohrian concept of complementarity is superseded by the much better defined notion of contextuality \cite{ITP2}.
%
\section{Sector parameter}
%
In order to describe the interaction between a microscopic quantum system and a macroscopic measurement device, we need a tool to label the state of the resulting {\color{black}joint} macroscopic system. We have shown in previous papers how the sectorisation theorems of von Neumann \cite{JvN39} can be used to understand the quantum states of macroscopic systems viewed as systems with a number $N\rightarrow\infty$ of microscopic quantum elements. {\color{black} Our considerations on sector parameters} are reminiscent of earlier works, e.g. by J.~Bub \cite{BubFoP88} and K.K.~Wan \cite{WanCJP80}, with the difference that we take explicitly care of the behaviour as a function of $N,$ which {\color{black} yields interesting physics. 
\\
\vskip -1mm
Let $\mathcal{S}$ be a system made of $N$ quantum subsystems, each described by an `elementary' Hilbert space $\mathcal{H}_\alpha, \alpha\in[N]:=\{0,1,...,N\}$. The Hilbert space of $\mathcal{S}$ is $\mathcal{H}_N := \otimes_{\alpha \in [N]}\mathcal{H}_\alpha$. 
For a collection of elementary states, $\{ \vert \phi_\alpha\rangle, \alpha \in [N]\}$ we can define a state of $\mathcal{S}$ in $\mathcal{H}_N$ as 
\begin{equation}
\vert \Phi_N\rangle = \otimes_{\alpha \in [N]} \vert \phi_\alpha\rangle 
\end{equation}  
In the large $N$ limit, such a state will define a von Neumann sector of $\mathcal{H}_\infty$, which is spanned by $\vert \Phi_\infty\rangle$ and all the states obtained by changing a finite number of $\vert \phi_ \alpha\rangle $'s in the tensor product \cite{ITP1,ITP2,ITP3,JvN39}. 
{\color{black}In turn, changing an infinite number of tensor factors will lead to a sector different from that of $\vert \Phi_\infty\rangle$}.

{\color{black}By analogy with the order parameter in statistical physics}, we define a `sector parameter' as the $N\rightarrow \infty$  limit of the observable associated to $\vert \Phi_N\rangle$ by 
\begin{equation}
\hat X_N:=\frac{1}{N}\sum_{\alpha \in [N]} \vert \phi_\alpha\rangle\langle\phi_\alpha\vert\otimes_{\beta\in[N]\setminus \alpha}\hat I_\beta
\end{equation}
where $\hat I_\beta$ is the identity operator in $\mathcal{H}_\beta$. 
This observable has the following properties: {\color{black}
\vskip 1mm
\noindent $\bullet$ $\vert \Phi_N\rangle$ is the eigenvector of $\hat X_N$ with eigenvalue 1:
\begin{equation}
\hat X_N \vert \Phi_N\rangle = \vert \Phi_N\rangle
\end{equation}
%
\noindent $\bullet$ The expectation value of $\hat X_N$ in a product state built by modifying $\vert \Phi_N\rangle$ on $M$ of its tensor factors departs from 1 by a quantity of order $M/N$. If for an $M$-element subset $\mathcal{C}$ of $\mathcal{S}$ (with indices in $I_\mathcal{C}\subset [N]$), the elementary states are $\vert\psi_\alpha \rangle$ instead of $\vert\phi_\alpha \rangle$, the state of $\mathcal{S}$ becomes
\begin{equation}
\vert \Psi_N\rangle := \otimes_{\alpha \in I_\mathcal{C}} \vert\psi_\alpha \rangle \otimes_{\beta \in [N]\setminus I_\mathcal{C}} \vert\phi_\beta \rangle
\end{equation}
then 
\begin{equation}
 \hat X_N \vert \Psi_N\rangle = 
 (\sum_{\alpha\in I_\mathcal{C}}   \frac{\langle \phi_\alpha\vert\psi_\alpha\rangle}{N})   \vert \Phi_N\rangle +
(1- \frac{M}{N} )  \vert \Psi_N\rangle
\end{equation}
\begin{equation}
\langle \Psi_N\vert \hat X_N \vert \Psi_N\rangle = 1 +\frac{1}{N}\sum_{\alpha\in I_\mathcal{C}} (\vert \langle \phi_\alpha\vert\psi_\alpha\rangle\vert^2-1)
\end{equation}
This means that (i) if $M$ remains finite when $N\rightarrow \infty$, then $X_N \vert \Psi_N\rangle \rightarrow \vert \Psi_N\rangle$, therefore all vectors of $\vert \Phi_\infty\rangle$'s sector are eigenvectors with the same eigenvalue 1;
and (ii)  if $\xi:=\vert \mathcal{C}\vert/N = M/N$ remains finite when $N\rightarrow \infty$, then 
$\langle \Psi_N\vert \hat X_N \vert \Psi_N\rangle -1$ has a limit of order  $\xi$.
In this sense, the value $\langle X_N\rangle$ distinguishes  the different sectors. 
\vskip 2mm
\noindent $\bullet$ The limiting value $\hat X_\infty :=\lim_{N\rightarrow\infty}\hat X_N$ is in the centre of the type-III algebra\footnote{{\color{black} The centre of an algebra is the set of operators that commute with all others}} 
of the ITP operators acting on $\mathcal{H}_\infty$. As a matter of facts, $\mathcal{H}_\infty$ is the direct sum of 
all the sectors. 
In each sector, $\hat X_\infty$ is proportional to the  sector's identity since the above (i) shows that there is a basis of the sector where this is the case. As far as inter-sector terms are concerned, they vanish thanks to the sectorisation theorem. 
\vskip 2mm
\noindent $\bullet$ Considering a second sector parameter $ \hat X^\prime_N$ built with a different set of elementary states $\{ \vert \phi^\prime_\alpha\rangle, \alpha \in [N]\}$ with\footnote{\color{black} This can be seen as corresponding to polarisation measurements along a different angle} 
$0 < \vert \langle \phi^\prime_\alpha \vert \phi_\alpha \rangle \vert < 1$, one can show that $[\hat X_N,\hat X^\prime_N] \sim 1/N$. Therefore the limit of the two sector parameters are two different, non-homothetic elements of the centre of the operator algebra on $\mathcal{H}_\infty$. This algebra is thus richer than a usual von Neumann factor, where the center is only the identity, up to a scalar.
}
\vskip 2mm
So overall, $\hat X_\infty$ is non-trivially diagonal, with a diagonal value 
$\lim_{N\rightarrow\infty}\langle \Phi_N\vert \hat X_N \vert \Phi_N\rangle$ in each sector defined by 
{\color{black} a corresponding ITP  $\vert \Phi_\infty\rangle$ as introduced above. }
%
\section{Modeling an avalanche photodiode}
Our goal is to 
spell out how sectorisation is at work in a measurement process, 
and to show that the breakdown of the Hilbert space at the $N\rightarrow\infty$ limit has physically meaningful 
{\color{black}precursors} 
before reaching this limit. This means that -- at least in this case -- the limit is under control and can be 
{\color{black} trusted} as part of the model. 
We take {\color{black}inspiration} for this {\color{black}on} the example of a standard photon polarisation measurement. 
{\color{black} Note however that as compared to the usual models (that involve Zeh-Zurek decoherence, see e.g. \cite{zurek}), we simplify deeply the description {\color{black} to keep track of the states' coherence}  as long as possible. This allows showing that during the amplification stage of the measurement -- that all devices involve at some point -- sectorisation alone could provide the properties of measurement. In this way, our model is somewhat more universal than the polarisation case we study.} 

The setup is described in Fig.~1. 
An incoming photon in state $\vert \gamma \rangle = h\vert H \rangle +v \vert V\rangle$  reaches a polarising beamsplitter (PBS),
where $V$ resp. $H$ stand for the vertical resp. horizontal polarisation in the PBS reference frame, and each output port
is connected to an avalanche photodiode (APD). 
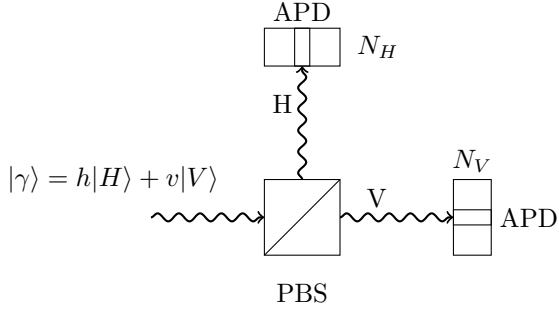
\begin{figure}
\centering
\begin{tikzpicture}

  \draw[->, thick, decorate, decoration={snake, amplitude=1.5pt, segment length=8pt}] (-2,0) -- (-0.5,0)
    node at (-2.5,0.5) {$\vert \gamma \rangle = h\vert H \rangle +v \vert V\rangle$};
  \draw[->, thick, decorate, decoration={snake, amplitude=1.5pt, segment length=8pt}] (0,0.5) -- (0,2)
    node[midway, above left] {H};
  \draw[->, thick, decorate, decoration={snake, amplitude=1.5pt, segment length=8pt}] (0.5,0) -- (2,0)
    node[midway, above left] {V};

  \draw (-0.5,2.0) rectangle (0.5,2.5);
  \draw (-0.1,2.0) rectangle (0.1,2.5);
  \node at (0,2.71) {APD};
  \node at (1,2.25) {$N_H$};

  \draw (2,-0.5) rectangle (2.5,0.5);
  \draw (2,-0.1) rectangle (2.5,0.1);  
  \node at (3.,0) {APD};
  \node at (2.25,0.75) {$N_V$};

  \draw (-0.5,-0.5) rectangle (0.5,0.5);
  \draw (-0.5,-0.5) -- (0.5,0.5);
  \node at (0,-1) {PBS};

\end{tikzpicture}
\label{fig:polMeas}
\caption{Example -- here $S$ is a photon in state $\vert \gamma\rangle$ and $M$ is a polarisation beamsplitter and  two photodiodes. 
In either APD, an avalanche that involves $N_P$ electron-hole pairs ($P=V,H$) might occur. 
These two cases correspond to two different sectors of $M$, so one can define a sector parameter as linked to the state of the APDs.
} 
\end{figure}
The photodiode is a piece of semiconductor of width $a$ doped with {\color{black}$A$} impurities that have energy levels $\vert\bot_n\rangle$ {\color{black}($n\in [A]$)} in the semiconductor gap at an energy $\Delta$ below the conduction band. The semiconductor is polarised by a potential $U$. When a photon of energy $\hbar \omega=\Delta$ is absorbed by impurity $n$, it excites an electron in a conduction-band state $\vert\top_n\rangle$ with  $\delta$ probability amplitude, related to the initial detection efficiency. 
\begin{equation}
\vert 1_\gamma\rangle \otimes \vert \bot_n\rangle \rightarrow \delta \; \vert 0_\gamma \rangle\otimes\vert \top_n\rangle + \sqrt{1-\vert \delta\vert ^2}\; \vert 1_\gamma \rangle\otimes\vert \bot_n\rangle 
\label{photexcitation}
\end{equation}
{\color{black} Note that we do not consider the conduction band as a continuum -- after all, the diode width can be seen as finite -- which means that Fermi's golden rule does not apply, and that no decoherence steps in yet.}
This electron is then accelerated in the conduction band, and when it has acquired an extra energy $\Delta$, it is likely to collide with a second impurity and excite a second electron {\color{black}into} the conduction band, while remaining there too. This triggers an avalanche of excited electrons that leads to the macroscopic measurable signal.
   {\color{black} The mean free path $l$ of the excited electrons can be estimated as $l/a=\Delta/(Ue)$, where $e$ is the electron's charge and it ends when the width of the semiconductor has been reached. 
This leads to approximately $g \sim Ue/\Delta$ generations of excited electrons, so $M=2^{Ue/\Delta}$ electrons.} This  is typically a very large number that saturates the available $A$. {\color{black} The work $W_m$ needed to perform the measurement is thus of the order of  $W_m \sim \Delta \min\{A,2^{Ue/\Delta}\}$, at a macroscopic scale.}
{\color{black} Note here again that we neither consider here interaction with a phonon bath in the semiconductor cristal nor resistivity in the wires that could bring more decoherence.}
The APD on the output port $P (=V, \, H)$ is assumed to be initially in a non-excited state
$$\vert \Omega^P_{[A]} \rangle := \otimes_{n\in[A]} \vert \bot_n^P\rangle$$
and the state of this APD {\color{black} after the $n$-th generation of the avalanche} will be noted $\vert \Phi^P_n\rangle$, to be detailed {\color{black}in the next section; we will show {\color{black} that }it depends critically on the efficiency of the avalanche}. The measurement process sequence thus starts from an initial state
\begin{equation}
\vert \Psi_{in} \rangle := \vert \gamma \rangle \otimes \vert \Omega^V_{[A]} \rangle \otimes \vert \Omega^H_{[A]} \rangle,
\end{equation}
then the photoexcitation leads to a first state
\begin{align}
\vert \Psi_0\rangle = &\sqrt{1-\vert \delta\vert^2}  \; \vert \gamma\rangle \otimes \vert \Omega_{[A]}^H \rangle \otimes \vert \Omega_{[A]}^V \rangle + \\ \nonumber
&\delta  \; \vert 0_\gamma\rangle \otimes(h\vert \Phi_0^H\rangle \otimes \vert \Omega_{[A]}^V \rangle + v \vert \Phi_0^V\rangle\otimes \vert \Omega_{[A]}^H \rangle)
\end{align}
{ \color{black} where, defining $[n:A]:=\{n,...,A\}$ for $n \geq 1$, one has}
$$
\vert \Phi_0^P\rangle:=\vert \top_0^P\rangle \otimes \vert \Omega_{[1:A]}^P \rangle.  
$$
After the $n$-th generation of collisions that involves $M=2^n$ electrons, the state writes
\begin{align}
\vert \Psi_n\rangle = &\sqrt{1-\vert \delta\vert^2} \; 
\vert \gamma\rangle \otimes \vert \Omega_{[A]}^H \rangle \otimes \vert \Omega_{[A]}^V \rangle \nonumber  +\\ 
&\delta  \; \vert 0_\gamma\rangle \otimes(h\vert \Phi_n^H\rangle \otimes \vert \Omega_{[A]}^V \rangle 
+ v \vert \Phi_n^V\rangle \otimes \vert \Omega_{[A]}^H \rangle).
\end{align}
{\color{black} The key role of our simplifying assumptions is {\color{black}to} allow us having a many-body pure state up to this stage.}

As we are dealing with polarisation measurements, the sector parameter can be defined with two macroscopic values $(+1,-1)$, that correspond to two sectors, `avalanche in channel $H$' or `avalanche in channel $V$'. It writes for $M=2^n$ electrons in the avalanche as
\begin{eqnarray}
\hat P_M :=&&\vert 0_\gamma\rangle \otimes|\Phi^H_n\rangle \otimes |\Omega^V_{[A]}\rangle \langle 0_\gamma\vert \otimes\langle\Phi^H_n | \otimes \langle\Omega^V_{[A]}| - \nonumber \\
&&\vert 0_\gamma\rangle \otimes|\Phi^V_n\rangle \otimes |\Omega^H_{[A]}\rangle \langle 0_\gamma\vert \otimes\langle\Phi^V_n |  \otimes \langle\Omega^H_{[A]}|  
\end{eqnarray}
{\color{black} Let us note that the APD is a strongly correlated quantum system, where the avalanche creates a strongly entangled state between many electrons; but ultimately only the value of the sector parameter (telling on which side the click happened) is of physical relevance.}
\section{Properties of the avalanche state}
\subsection{The `avalanche state'}
To exploit the above considerations, we need to explore the properties of $\vert \Phi^P_n\rangle$.
Let us assume that the collision process triggered
with amplitude $\eta$ by conduction-band electron $j$ on dopant-impurity electron $k$ writes 
\begin{equation}
\vert\top_j \rangle \otimes \vert \bot_{k}\rangle \rightarrow \vert\top_j \rangle \otimes (\eta \vert \top_{k}\rangle + \sqrt{1-\vert \eta \vert^2}\vert \bot_{k}\rangle)  
\label{excitation}
\end{equation}
{\color{black}
Electron after electron, this triggers an avalanche of collisions. 
 At generation $n$, the resulting state of the APD on output $P$ with an avalanche that involves $M=2^n$ electrons writes (see Annex 3) 
\begin{eqnarray}
&&\vert \Phi^P_{n+1} \rangle = \nonumber \\
&&\vert Z^P_0(i^{n+1}_0) \rangle \otimes \vert Z^P_1(i^{n+1}_1) \rangle \otimes \vert Z^P_2(i^{n+1}_2,i^{n+1}_3) \rangle \otimes ...  \nonumber \\
&&\otimes \vert Z^P_{n+1}([i^{n+1}_{2^n-1}:i^{n+1}_{2^{{n+1}}-1}]) \rangle \otimes \vert\Omega^P([i^{n+1}_{2^{n+1}}:A)\rangle\nonumber
\end{eqnarray}
with avalanche efficiency taken care of by
\begin{eqnarray}
&& \vert Z^P_k([i^n_{2^{k-1}}:i^n_{2^{k}-1}]) \rangle = 
\sqrt{1-\vert \eta \vert^2}\vert \Omega^P([i^n_{2^{k-1}}:i^n_{2^{k}-1}]) \rangle  \nonumber \\
&&+...+ \eta \otimes_{l=0}^{n-1}\vert Z^P_l([i^n_0:i^n_{2^{l-1}-1}])\rangle
\end{eqnarray} 
where the sets of electron indices $(i^n_0,...,i^n_{2^k-1})$ are disjoint partitions of $[2^n]$ for different $k$ values, and   
$$
\vert Z_0^P\rangle = \vert \top_0^P\rangle.
$$
Note that as mappings of eqs.~(\ref{photexcitation}) and (\ref{excitation}) are unitary, there is no loss of quantum information in this process.
}
\subsection{Properties}
{\color{black}
Before going to the specific polarisation measurement results, it is interesting to investigate the large $N$ relative properties of the two would-be sector-reference states $ \vert \Phi_n^P\rangle$ and $ \vert \Omega_{[A]}^P \rangle$. One can show {\color{black}(see Annex 3)} that at leading order 
\begin{equation}
\langle \Omega_{[A]}^P \vert \Phi_n^P\rangle \sim (\sqrt{1-\vert \eta\vert^2})^{n-1} \rightarrow 0
\label{orthog}
\end{equation}
This property is of high interest if considered in the frame of the generic model for quantum measurements of \cite{genModel}. That model gives the measurement outcome probability depending on the type of ancilla or meter states involved in the measurement process. Here, $ \vert \Phi_n^P\rangle$ and $ \vert \Omega_{[A]}^P \rangle$ are meter states that result from coupling the APD with the system under measurement. When the number of electrons involved in the avalanche is small, these states are not orthogonal. This corresponds to a reversible situation where  interferences terms would be  needed to compute measurement-outcome probability. When $n$ grows, these states become more and more orthogonal, which corresponds to the case where interferences disappear and probabilities, not amplitudes, are added to compute the measurement-outcome probabilities. In other words, the avalanche drives the phenomenon at stake from a reversible to an irreversible situation, and thus to the measurement outcome. 

This is obtained simply from the properties of the Hilbert space, and not by tracing out on external degrees of freedom, and happens rather {\color{black} gradually} despite the qualitative change that occurs at the limit. 
It is of further interest to note that in this measurement model, for a system with $D$ possible measurement outcomes, this property is needed for only $D-1$ of the meter states 
associated with different outcomes, allowing ``excluded-middle measurements'' to be possible~\cite{whitaker}.}
\subsection{Measurement}
Let us now consider the full measurement setup with both photodiodes. After generation $n$, the density operator writes 
\begin{align}
\hat \rho_n &:= \vert \Psi_n \rangle \langle \Psi_n\vert  \\
& = (1-\vert \delta\vert^2) 
\vert \gamma\rangle  \vert \Omega_{[A]}^H \rangle  \vert \Omega_{[A]}^V \rangle \langle \gamma\vert  \langle \Omega_{[A]}^H \vert  \langle \Omega_{[A]}^V \vert \nonumber \\
&+ \vert h\delta  \vert^2 
\vert 0_\gamma\rangle \vert \Phi_n^H\rangle  \vert \Omega_{[A]}^V\rangle
\langle 0_\gamma\vert \langle \Phi_n^H\vert  \langle \Omega_{[A]}^V \vert \nonumber \\
&+ \vert v\delta  \vert^2 \vert 0_\gamma\rangle\vert \Phi_n^V\rangle \vert \Omega_{[A]}^H \rangle
\langle 0_\gamma\vert\langle \Phi_n^V\vert \langle \Omega_{[A]}^H \vert\nonumber \\
&+\delta^* h^* \sqrt{1-\vert \delta\vert^2} \; \vert \gamma\rangle \vert \Omega_{[A]}^H \rangle  \vert \Omega_{[A]}^V \rangle \langle 0_\gamma\vert \langle \Phi_n^H\vert  \langle \Omega_{[A]}^V \vert +h.c.\nonumber \\
&+\delta^* v^* \sqrt{1-\vert \delta\vert^2}  \; \vert \gamma\rangle \vert \Omega_{[A]}^V \rangle  \vert \Omega_{[A]}^H \rangle \langle 0_\gamma\vert \langle \Phi_n^V\vert  \langle \Omega_{[A]}^H \vert +h.c.\nonumber \\
&+\vert \delta\vert^2 hv^*\vert 0_\gamma\rangle \vert \Phi_n^H\rangle  \vert \Omega_{[A]}^V\rangle
\langle 0_\gamma \vert \langle \Phi_n^V\vert \langle \Omega_{[A]}^H \vert +h.c.\nonumber
\end{align}
With this description of the avalanche state, one can compute the expectation value of the above-defined sector parameter as the avalanche unfolds. 
\begin{equation}
\langle \hat P_{\color{black}2^n}\rangle := \mathrm{Tr}(\hat \rho_n  \hat P_{\color{black}2^n}) = \langle\Psi_n| \hat P_{\color{black}2^n} |\Psi_n\rangle 
\end{equation}
which after some elementary algebra yields
\begin{eqnarray}
&\langle \hat P_{\color{black}2^n}\rangle
= |\delta|^2  (|h|^2 - |v|^2) (1-   |\langle  \Phi_n^H  \vert  \Omega^H_{[A]}\rangle \langle  \Phi_n^V  \vert  \Omega^V_{[A]}\rangle |^2)
\nonumber \\
\label{yield}
\end{eqnarray}
which is produced by $H-H$ or $V-V$ diagonal blocks in  $\hat \rho_n$.
 As a result of eq. (\ref{orthog}) 
 the expectation value of $\hat P_{\color{black}2^n}$ converges to the expected 
$\vert\delta\vert^2 (\vert h\vert^2-\vert v\vert^2)$ when the avalanche heads towards the sectorisation limit.
This means that although it requires taking the limit to reach the complete sectorised Hilbert space, early signs of sectorisation exist before as qualitative changes set in continuously. 

\section{Discussion}
\subsection{Summary of the argument}
{\color{black}
In previous papers, we have argued that the sectorisation occurring in an infinite tensor product of Hilbert space could shed light on the relationship between the quantum and the classical world. In this paper, we have addressed the question of the validity of taking this infinite limit by investigating how a measurement setup, which is typically a device that connects the quantum and the classical, can be described as a physical implementation of the path towards this ITP limit. This setup  performs a photon polarisation measurement and involves two APDs. We have explored the large particle number limit of the avalanche pure state in the APDs.}

We have shown {\color{black} that} even though our model does not involve usual sources of decoherence (continuum of the conduction band, phonon thermal bath, resistivity, etc -- no Lindblad equation is invoked either) before the measurement signal gets amplified, the measured quantities are the ones obtained with standard decoherence. Here, decoherence results from the sectorisation of the Hilbert space with the {\color{black} double} exponential divergence of its size. More precisely, for a large number of  particules, all happens as if there were no inter-sector contribution to measurable quantities. This means that even before reaching the $N\rightarrow \infty$  limit, the sectorisation behaviour sets in {\color{black} and converts the pure state in an effective mixed state. This limit being 
regular, {\color{black} it can be considered as {\color{black}as} legitimate as taking the thermodynamic limit in Statistical Physics.
%
\subsection{Generalised picture}
{\color{black} This measurement scenario can be summarised in a language that can be generalised in the following way, 
using some vocabulary spelled out in Annex 1. 

\noindent $\bullet$ Before measurement, system and context are well separated and do not interact. The system is prepared in one of the modalities of context $\mathcal{C}$ and its observables are described by operators in a type-I $W^*$-algebra. The measurement device that defines the context is in a well defined sector ($\vert \Omega^H_\infty, \Omega^V_\infty\rangle$), and its observables are operators in a type-III $W^*$-algebra. 

\noindent $\bullet$ The state analyser (here the PBS) is present and defines a new context $\mathcal{C}^\prime$ for the measurement, but has no effect before the system reaches a detector.

\noindent $\bullet$ In the detector (here the APD), interaction starts, and the system gets entangled with an exponentially increasing number of electrons taken from the context. This chain reaction amplifying the photodetection creates a bigger and bigger system, but no measurement result yet,
{\color{black} it is like having a larger system.}

\noindent $\bullet$  At some point, the number of electrons fed into the avalanche is so large that it results in a macroscopic change that cannot be ignored. This is visible in the computation of physical quantities that converge towards those obtained with a mixed state despite the avalanche being described by a pure state. The polarisation is no more defined along one of the directions of context $\mathcal{C}$ but rather along one of context $\mathcal{C}^\prime$.

\noindent $\bullet$ The state analyser is already oriented along the directions of $\mathcal{C}^\prime$ before the system reaches the detector, and its role is to structure the upcoming macroscopic effects, by defining which set of modalities can be amplified -- in other words, it choses along which of its subspaces the divergent Hilbert space will be broken down into an effective mixed state. Only later, when the system reaches the detector, the measurement result is actualised by the avalanche. This is the case even in the case of an excluded-middle measurement \cite{whitaker}, where only $D-1$ detectors are positioned, one of the measurement issues being potentially unread -- once the context is defined, detectors for $D-1$ orthogonal meter states are enough to actualise the measurement.}

\noindent $\bullet$ The situation of a destructive measurement considered above can be easily extended to a QND measurement, see Annex 2. 
\subsection{Some previous work}
Other attempts at describing measurement processes with sectorisation have been done over the past decades, but have received moderate support \cite{opus}. Hepp \cite{Hepp} developed a model that was criticised by Bell \cite{BellOnHepp} as irrealistic because it required an infinite time to converge {\color{black} (contrary to the APD, in Hepp's model, the size of the Hilbert space grows linearly, not exponentially with time)}. Emch \cite{EmchMeasurement}, Araki \cite{ArakiMeasurement} and Bub \cite{BubFooP} developed models that assumed sectorisation was in place from the beginning instead of building up through a dynamical process. This yielded interesting results at the limit, but left open the question of the validity of the limit itself. Our approach is addressing simultaneously the acceleration of the convergence to the limit via the exponential avalanche, and the quantitative way this convergence occurs on physical quantities, thus validating the limit. 
\vskip 2 mm

Ellis and Drossel have done a thorough analysis of the same measurement process \cite{DrosselEllis}. They propose a scenario  through successive steps where decoherence is not due to the inherent sectorisation of the Hilbert space, but rather to external thermal baths, modeled with Lindblad equations. Even if we show that the thermal baths are not necessary (which does not prevent a fortiori from adding them to sectorisation effects), the steps they propose can still be used in our view to get an insight on the process, as they are provided by the amplification step {\color{black} with the following changes}: 
\vskip 1mm
\noindent $\bullet$ {\it Divergence of the dimension of the Hilbert space} of (system + involved part of the instrument) that 
splits into orthogonal sectors, driven by the design of device (vector sectorisation theorem). The sectors are labelled by the values of the sector parameter and correspond to the measured quantity. This design fixes a measurement context, and yields the modalities (see Annex 1) that can be actualised by the setup.
\vskip 2 mm

\noindent $\bullet$ {\it Trapping of the state} of (system + instrument) in one 
sector, since transitions between sectors become more and more unlikely as dimension grows (operator sectorisation theorem). Deciding the destination sector cannot be done deterministically because the information content of the system state is finite, while it would take a much larger amount of information to specify the state of all the elementary parts of the sector state.
\vskip 1mm
\noindent $\bullet$ {\it Amplification} itself, that allows a value readout. 
As a conclusion about this comparison, adding more decoherence effects does not harm and does not contradict our approach, which provides a self-decoherence mechanism that is able to physically and mathematically ``terminate'' the measurement process. 
This last step may not be obvious to achieve by using only the standard decoherence approach. 
\subsection{Concluding remarks}
To conclude, a few remarks are in order.
First, in this scenario, we note that one always needs the macroscopic cascade to be fed by energy, and in our case, the avalanche is fed by the polarisation potential of the diode. This potential is treated classically, and (at the limit) plays the role of an infinite resource that makes the process possible, within a macroscopic context{\color{black}, set by the value of the measurement work $W_m$. This is consistent with the considerations of \cite{InfResource}, even though in our case temperature is not involved and the third principle of Thermodynamics not at play.}

{\color{black} Second, in the case where $\eta= 1$, the two states that correspond to avalanches on either sides of the polarisation beamsplitter are directly orthogonal.  Their superposition due to the initial splitting of the photon is thus a so-called ``cat state’’, and the exponentially small terms in the  sector parameter (\ref{yield})  are directly zero, so the cat is either living or dead. Obviously this does not prevent other more tricky observables to reveal possible interferences between the two (finite) branches \cite{BellOnHepp,bori}; however such observables are guaranteed to vanish at infinity, thanks to the sectorisation theorem. Let us emphasize again that in our approach, and for any $\eta$,  the final situation with  a new result in a new context (a new modality, using the terminology of Annex 1) does not ``emerge’’, but is warranted by the whole construction. Correspondingly, the algebraic construction  determines the asymptotic  modalities, either long after or long before the measurement itself; whereas what happens ``during the measurement’’ is described in an approximate way, but with known boundaries. }
\vskip 3mm

Third, and quite importantly, this view allows us to
narrow down the location of the Heisenberg cut, that can be traced to a change in the algebraic properties of the diverging-size Hilbert space. The cut lies where it is no more possible to make an experimental difference between a separable and a non-separable Hilbert space{\color{black}~--~said otherwise, when it is no more possible to tell whether the Hilbertian basis is countable or not. It would be desirable to find more quantitative criteria based on this idea,  this is left for further work.}
\vskip 2mm

{\color{black} Finally, one}  further sees that there is again no clear bottom-up or top-down causation of the behaviours -- the microscopic, quantum, realm needs the 
macroscopic world to manifest its properties, and the macroscopic world could not exist without its microscopic quantum elements. This fits quite well with the CSM approach (see Annex 1) but clearly differs from traditional views looking for an ``emergence of the classical''. 
\\

{\bf Acknowledgements.} The authors thank Daniel Est\`eve, 
Alexia Auff\`eves, Jeffrey Bub, Borijove Daki\'c, 
Franck  Lalo\"e and Roger Balian for many discussions.

\section*{\large Annex 1}
Here we write a few words about the general framework we use, that does not fit into any usual interpretations of quantum mechanics; it is quoted as CSM (Contexts, Systems, Modalities), and to put it in a quantum foundations box it might be called neo-Copenhagian, where ‘neo’ is more important than ‘Copenhagian’. In particular, we don't invoke Bohr's complementarity, but we give a central role to contextuality, and to the realist ontology called Contextual Objectivity \cite{CO2002}. 

The basic idea is that physical objects are (quantum) systems within (classical) contexts, and they carry real (certain and repeatable) physical properties called modalities \cite{csm1}. A usual state vector is a mathematical object attached to an equivalence class of modalities, that are mutually certain though belonging to different contexts; this equivalence relation of modalities is called extravalence \cite{csm2}. A closely related concept is called intertwinning (of contexts), and appears in Gleason's theorem \cite{csm4}. These definitions have the big advantage to provide a clear distinction between physical objects (systems within contexts, carrying modalities) and  mathematical objects (usual vector states $|\psi \rangle$), that are tools to calculate probabilities of transition between modalities. Probabilities are provably required because modalities are quantized (the number of mutually exclusive modalities depends on the system, but not on the context), and contextual (by construction, as written above); then Born’s rule can be demonstrated from Uhlhorn’s and Gleason’s theorems \cite{csm4}.

Due to the postulated existence of (quantum) systems within (classical) contexts, the Heisenberg cut is built in the theory from the beginning, and it can be recovered at the end by using operator algebra and infinite tensor products (ITP) \cite{completing,ITP1,ITP2,ITP3}.
In the present paper we show explicitly how classical-looking quantities appear as ‘sector parameters’ during a quantum measurement; remember this is not an ‘emergence’, but a consistency check, since both systems and contexts are already there at the starting point of the construction. This echoes in a positive sense Lev Landau’s famous sentence, {\it ‘Quantum mechanics occupies a very unusual place among physical theories: it contains classical mechanics as a limiting case, yet at the same time it requires this limiting case for its own formulation’. }

On the formal side, our point of view is close to the one expressed for instance by Jeff  Bub in \cite{BubFoP88,BubFooP,bub}. However Bub’s information-theoretic interpretation necessarily rises the question of ‘information about what?’ CSM answers this question, since it provides on the physical side an ontology based on contextual objectivity \cite{CO2002}, corresponding on the mathematical side to a formalism based on operator algebras. 

Finally, we note that taking infinite limits has been heavily criticized, e.g. in \cite{earman} by John Earman, or in \cite{BellOnHepp} by John Bell  who writes: 
{\it ``The continuing dispute about quantum measurement theory is (...) between people who view with different degrees of concern or complacency the following fact: so long as the wave packet reduction is an essential component, and so long as we do not know exactly when and how it takes over from the Schr\"odinger equation, we do not have an exact and unambiguous formulation of our most fundamental physical theory.''} 
Our conclusion from this sentence is that the problem has an ontological origin, that is: what are the physical objects?  what can we expect from a physical theory ? Then one should remember that our goal here is not to have the classical world ``emerging'' from the quantum one in a reductionist approach: both of them are already there from the initial postulates in the CSM framework. So what has to be established is the consistency of the overall picture, both a physical and a mathematical point of view, see for instance  the discussion in Section III of \cite{ITP3}.

\section*{\large Annex 2}
In the main text, we consider an incoming photon in state $\vert \gamma \rangle = h\vert H \rangle +v \vert V\rangle$, which  reaches a polarising beamsplitter PBS,
where $V$ or $H$ stand for the vertical or horizontal polarisation, and each output port
is connected to an avalanche photodiode (APD). This is called a  direct, or destructive measurement, where the incoming photon disappears. A more interesting scheme is a Quantum Non-Demolition (QND) measurement, where the system is left in the measured state. This can be done by using an ancilla photon, such as the photon $s$ to be measured, initially in state  $\vert \gamma_s \rangle = h\vert H_s \rangle +v \vert V_s\rangle$, is entangled with the ancilla photon $m$ to produce the state $\vert \gamma_{sm} \rangle = h\vert H_s H_m \rangle +v \vert V_s V_m \rangle$. In principle this can be achieved by performing a C-NOT gate  between the two photons, where the initial ancilla state $\vert \gamma_e \rangle = \vert H_e \rangle$ remains the same if $\vert \gamma_s \rangle = \vert H_s \rangle$, and is changed to  $\vert \gamma_e \rangle = \vert V_e \rangle$ if $\vert \gamma_s \rangle = \vert V_s \rangle$. Such gates are difficult to realize deterministically in the optical domain, but they are possible in principle using for instance Rydberg superatoms \cite{ao}. 
Then the previous scheme using a polarising beamsplitter PBS and two APDs can be used with the ancilla photon, and from the usual properties of the state $\vert \gamma_{sm} \rangle$ the photon $s$ is left in state $\vert H_s \rangle$ with probability $|h|^2$, and in state $\vert V_s \rangle$ with probability $|v|^2$, as expected. We note that physically equivalent schemes have been implemented with two trapped ions and irreversible photodetection, see a full discussion in \cite{ITP1}.  
\\

Such a scheme is quite generic, and the $s$-$m$ entanglement step is often called a pre-measurement;  obviously it is not conclusive as long as the avalanche and sectorisation have not happened on the ancilla. The fact that  conclusive measurements do happen in a single macroscopic universe is a distinctive feature of the CSM approach  \cite{ITP1,ITP2,ITP3}. 

\section*{\large Annex 3}
\subsection*{Form of the `avalanche state'}
As the avalanche unfolds, we assume the photodiode goes through a succession of pure states that involve more and more excited electrons. In this model, we assume that there are $A$ available electrons in dopant impurities in the semiconductor gap. For electron $k$, we note $\vert \bot_k \rangle$ its state when it is located in its impurity and $\vert \top_k \rangle$ its state when it is excited in the conduction band. We note 
$\vert \Omega(i,j,...,n) \rangle := \vert \bot_i \rangle\otimes \vert \bot_j \rangle \otimes ...\otimes  \vert \bot_n \rangle$ 
Right after the absorption of the photon, the state is 
$$
\vert \Phi_0\rangle = \vert \top_0 \rangle \otimes \vert \Omega([1:A) \rangle
$$
The scattering towards the conduction band of an impurity electron $j$ by a conduction band electron $i$ with amplitude $\eta$ is described by an operator $\hat S_{i,j}$ such that 
\begin{eqnarray}
\hat S_{i,j} \vert \top_i \rangle \otimes \vert \bot_j \rangle &&= \vert \top_i \rangle \otimes (\sqrt{1-\vert\eta|^2}\vert \bot_j \rangle + \eta \vert \top_j \rangle) \nonumber \\
\hat S_{i,j} \vert \bot_i \rangle \otimes \vert \bot_j \rangle &&= \vert \bot_i \rangle \otimes \vert \bot_j \rangle
\end{eqnarray}

\noindent In this simplified model, the first generation of excitation is described by 
$$
\vert \Phi_1\rangle = \hat S_{0,1} \vert \Phi_0\rangle = \vert Z_0(0) \rangle \otimes \vert Z_1(1)\rangle \otimes \vert \Omega([2:A]) \rangle
$$
with $\vert Z_0(i) \rangle := \vert \top_i \rangle$ and $\vert Z_1(j) \rangle := \sqrt{1-\vert\eta|^2}\vert \Omega(j) \rangle + \eta \vert Z_0(j) \rangle$. The second generation goes by 
\begin{eqnarray}
\vert \Phi_2\rangle =&& \hat S_{0,2} \hat S_{1,3} \vert \Phi_1\rangle \nonumber \\
=&& [\hat S_{0,2} \vert Z_0(0)\rangle\otimes\vert \Omega(2)\rangle]\otimes[\hat S_{1,3}\vert Z_1(1)\rangle\otimes\vert \Omega(3)\rangle]  \nonumber \\
&& \otimes \vert \Omega(4,...,A) \rangle  \\
=&& \vert Z_0(0) \rangle \otimes \vert Z_1(2)\rangle \otimes \vert Z_2(1,3)\rangle  \otimes \vert \Omega([4:A]) \rangle\nonumber
\end{eqnarray}
with $\vert Z_2(i,j) \rangle := \hat S_{i,j}\vert Z_1(i)\rangle\otimes\vert \Omega(j)\rangle =\sqrt{1-\vert\eta|^2}\vert \Omega(i,j)\rangle + \eta \vert Z_0(i) \rangle \otimes \vert Z_1(j) \rangle$. As for the third generation, the photodiode state reads
\begin{eqnarray}
\vert \Phi_3\rangle =&& \hat S_{0,4} \hat S_{1,5}\hat S_{2,6} \hat S_{3,7} \vert \Phi_2\rangle \nonumber \\
=&& [\hat S_{0,4} \vert Z_0(0)\rangle\otimes\vert \Omega(4)\rangle]\otimes[\hat S_{2,6}\vert Z_1(2)\rangle\otimes\vert \Omega(6)\rangle]  \nonumber \\
&& \otimes[\hat S_{1,5}\hat S_{3,7}\vert Z_2(1,3)\rangle\otimes\vert \Omega(5,7)]
\otimes \vert \Omega([8:A]) \rangle \nonumber \\
=&& \vert Z_0(0) \rangle \otimes \vert Z_1(4)\rangle \otimes \vert Z_2(2,6)\rangle \otimes \vert Z_3(1,3,5,7)\rangle\nonumber \\
 &&\otimes \vert \Omega([8:A]) \rangle
\end{eqnarray}
with $\vert Z_3(i,j,k,l) \rangle := \hat S_{i,j}\hat S_{k,l}\vert Z_2(i,k)\rangle\otimes\vert \Omega(j,l) 
= \sqrt{1-\vert\eta|^2}\vert \Omega(i,j,k,l)\rangle + \eta \vert Z_0(i) \rangle \otimes \vert Z_1(j) \rangle \otimes \vert Z_2(k,l) \rangle$. The grouping of terms according to the entangled subsets is illustrated in Fig.2.
\begin{figure}
\centering
\begin{tikzpicture}
  \draw (0,-2.4) -- (0,-1.5);
  \draw (0,-2.5) circle[radius=2pt];
  \draw (0,-1.5) -- (-0.5,-1);
  \fill (0, -1.5) circle[radius=2pt];
  \draw (-0.5,-1) -- (-1.25,-0.5);
  \fill (-0.5, -1) circle[radius=2pt];
  \draw (-1.25, -0.5) -- (-2.25, 0);
  \fill ((-1.25, -0.5) circle[radius=2pt];
  \node at (-2.25, 0.25) {0};
  \node at (-1.25, -2.5) {photoexcitation};
  \node at (1.25, -2.5) {$\vert Z_0(0)\rangle $};
  \draw (0.25, -1.5) -- (0.75, -1);
  \fill (0.25, -1.5) circle[radius=2pt];
  \draw (0.75,-1) -- (0.75, -0.5);
  \fill (0.75, -1) circle[radius=2pt];
  \draw (0.75,-0.5) -- (0.5,-0.);
  \fill (0.75, -0.5) circle[radius=2pt];
  \node at (0.5, 0.25) {1};
  \node at (1.75, -1.5) {$\vert Z_0(0)\rangle \vert Z_1(1)\rangle $};  
  \draw (-0.25, -1) -- (-0.5,-0.5);
  \fill (-0.25, -1) circle[radius=2pt];
  \draw (-0.5,-0.5) -- (-0.5, 0);
  \fill (-0.5,-0.5) circle[radius=2pt];
  \node at (-0.5, 0.25) {2};  
  \draw (1,-1) -- (1.5, -0.5);
  \fill (1,-1) circle[radius=2pt];
  \draw (1.5, -0.5) -- (1.75, 0);
  \fill (1.5, -0.5) circle[radius=2pt];
  \node at (1.75, 0.25) {5};
  \node at (3, -1) {$\vert Z_0(0)\rangle \vert Z_1(2)\rangle \vert Z_2(1,3)\rangle$};
  \draw (-1, -0.5) -- (-1.25, 0);
  \fill ((-1, -0.5) circle[radius=2pt];
  \node at (-1.25, 0.25) {4};
  \draw (-0.25,-0.5) -- (0,0);
  \fill ((-0.25,-0.5) circle[radius=2pt];
  \node at (0, 0.25) {6};
  \draw (1, -0.5) -- (1.25, 0);
  \fill  ((1, -0.5) circle[radius=2pt]; 
  \node at (1.25, 0.25) {3};
  \draw (1.75, -0.5) -- (2.5, 0);
  \fill (1.75, -0.5) circle[radius=2pt];
  \node at (2.5, 0.25) {7};
  \node at (2.5, 0.75) {$\vert Z_0(0)\rangle \vert Z_1(4)\rangle \vert Z_2(2,6)\rangle\vert Z_3(1,3,5,7)\rangle$};
  \node at (-0, 1.5) {...};
  \node at (-2, 1.5) {$t$};
  \draw[->] (-2.5, -2.5) -- (-2.5, 1.5);
  
  \node at (5, -1.5) [draw, rotate=90] {avalanche $\rightarrow$};

\end{tikzpicture}
\label{fig:apdCascade}
\caption{Labelling of the sequence of events following the absorption of a photon by a doping impurity in an APD. At generation $n$, electron $k\in[2^{n-1}-1]$ excites electron $k+2^{n-1}$ with the process of eq. (\ref{excitation}).
} 
\end{figure}

A regular structure thus appears, where 
\begin{eqnarray}
\label{eqnPhi}
&&\vert \Phi_{n+1} \rangle = \hat S_{0,2^n} \hat S_{1,1+2^n} ... \hat S_{2^n-1, 2^{n+1}-1} \vert \Phi_n\rangle  =  \nonumber \\
&& \vert Z_0(i^{n+1}_0) \rangle \otimes \vert Z_1(i^{n+1}_1) \rangle \otimes \vert Z_2(i^{n+1}_2,i^{n+1}_3) \rangle \otimes ...\nonumber \\
&&\otimes \vert Z_{n+1}([i^{n+1}_{2^n-1}:i^{n+1}_{2^{{n+1}}-1}]) \rangle \otimes \vert\Omega([i^{n+1}_{2^{n+1}}:A])\rangle
\end{eqnarray}
with
\begin{eqnarray}
&&\vert Z_k([i^n_{2^{k-1}}:i^n_{2^{k}-1}]) \rangle :=  \nonumber \\
&&\hat S_{i^n_{2^{k-1}},i^n_{2^{k-1}+2^{k-2}}} \hat S_{i^n_{2^{k-1}+1},i^n_{2^{k-1}+2^{k-2}+1}}...
\hat S_{i^n_{2^{k-1}+2^{k-2}-1},i^n_{2^{k}-1}} \nonumber \\
&&\vert Z_{k-1}([i^n_{2^{k-1}}:i^n_{2^{k-1}+2^{k-2}-1}])\rangle \otimes \vert \Omega([i^n_{2^{k-1}+2^{k-2}}:i^n_{2^k-1}])  \rangle \nonumber
\label{eqnForm}
\end{eqnarray}
and
\begin{eqnarray}
&&\label{eqnZ}
\vert Z_k([i^n_{2^{k-1}}:i^n_{2^{k}-1}]) \rangle = \sqrt{1-\vert \eta \vert^2}\vert \Omega([i^n_{2^{k-1}}:i^n_{2^{k}-1}]) \rangle \nonumber\\
&&...+ \eta \otimes_{l=0}^{n-1}\vert Z_l([i^n_0:i^n_{2^{l-1}-1}])\rangle
\end{eqnarray}
The latter can be proven by induction after, first conveniently grouping the operators according to the entangled subsets of electrons in equation (\ref{eqnPhi}) to get products of terms of the form of the right hand side of equation (\ref{eqnPhi}); second noting that the scattering operators $\hat S_{i,j}$ leave the pure $\vert \Omega\rangle$ states unchanged while, when they act on a previous $\vert Z_k\rangle$ state and an $\vert \Omega\rangle$ state, expanding and further distributing the indices in new $\vert Z_k\rangle$ states, thus producing order by order the form of eq. (\ref{eqnZ}).

\subsection*{{\color{black} Unitarity considerations}}

At this point, it can be noted that the scattering operator can be completed on the space where the second electron is already in the conduction band as 
\begin{eqnarray}
\hat S_{i,j} \vert \top_i \rangle \otimes \vert \top_j \rangle &&= \vert \top_i \rangle \otimes (\sqrt{1-\vert\eta|^2}\vert \top_j \rangle + \eta \vert \bot_j \rangle) \nonumber \\
\hat S_{i,j} \vert \bot_i \rangle \otimes \vert \top_j \rangle &&= \vert \bot_i \rangle \otimes \vert \bot_j \rangle
\end{eqnarray}
This completion does not change the above results, but it makes the scattering operators unitary, and thus the evolution along which the avalanche unfolds a unitary automorphism in the Hilbert space of the $A$ available electrons. {\color{black} It is however known that unitary equivalence breaks down at the infinite limit, and one can see the behaviours we exhibit as early signs of this breakdown}. 

\subsection*{Avalanche to no-avalanche overlap}
Let $\vert \Omega^\prime\rangle := \vert \top_0\rangle \otimes \vert \Omega(1,...,A)\rangle$ be the state where no avalanche occurs. It is of interest to compute $\langle \Omega^\prime\vert \Phi_n \rangle$.
$$\langle \Omega^\prime\vert \Phi_n \rangle = \prod_{l=1}^{n-1} \langle \Omega(2^{l-1},...,2^l-1)\vert Z_l(2^{l-1},...,2^l-1)\rangle.$$
Now, 
\begin{eqnarray}
&&\langle \Omega(2^{l-1},...,2^l-1)\vert Z_l(2^{l-1},...,2^l-1)\rangle \nonumber \\
&&= \langle \Omega(2^{l-1},...,2^l-1)\vert(\sqrt{1-\vert \eta\vert^2}\vert \Omega(2^{l-1},...,2^l-1) \nonumber \\
&&+\eta\otimes_{k=0}^{l-1}\vert Z_k(2^{k-1},...,2^k-1)\rangle)\nonumber \\
&& = o(\sqrt{1-\vert \eta\vert^2})
\end{eqnarray}
so 
$\langle \Omega^\prime\vert \Phi_n \rangle \sim (\sqrt{1-\vert \eta\vert^2})^{n-1}.$
Thus the avalanche state becomes gradually orthogonal to the no-avalanche state as the avalanche unfolds with increasing $n$.

\end{document}